\documentclass{PoS}
\usepackage{amsmath}

\newcommand{\vcb}{|V_{cb}|}
\newcommand{\vub}{|V_{ub}|}

\newcommand{\tev}{\, \text{TeV}}
\newcommand{\gev}{\, \text{GeV}}

\newcommand{\lsim}{
\mathrel{\hbox{\rlap{\hbox{\lower4pt\hbox{$\sim$}}}\hbox{$<$}}}}

\newcommand{\gsim}{
\mathrel{\hbox{\rlap{\hbox{\lower4pt\hbox{$\sim$}}}\hbox{$>$}}}}

\def\kpn{K^+\rightarrow\pi^+\nu\bar\nu}
\def\klpn{K_{L}\rightarrow\pi^0\nu\bar\nu}

\def\eps{\varepsilon}
\def\epe{\varepsilon'/\varepsilon}

\title{Impact of Lattice QCD on CKM Phenomenology}

\ShortTitle{Impact of Lattice QCD on CKM Phenomenology}

\author{\speaker{Monika Blanke}\\
        {Institut fur Kernphysik, Karlsruhe Institute of Technology,
  Hermann-von-Helmholtz-Platz 1,
  D-76344 Eggenstein-Leopoldshafen, Germany}\\
 {Institut fur Theoretische Teilchenphysik,
  Karlsruhe Institute of Technology, Engesserstra\ss e 7,
  D-76128 Karlsruhe, Germany} \\
        E-mail: \email{monika.blanke@kit.edu}}

\abstract{Precise lattice QCD results for hadronic matrix elements, decay constants and form factors play a crucial role in the determination of CKM matrix elements and in the identification of possible new physics contributions to flavour violating observables. This article reviews the implications of recent lattice QCD results on the phenomenology of flavour and CP violating meson decays, and highlights some future directions for lattice QCD calculations which would have a major impact on flavour phenomenology.}

\FullConference{34th annual International Symposium on Lattice Field Theory\\
		24-30 July 2016\\
		University of Southampton, UK}

\begin{document}

\section{Introduction}

With the discovery of the Higgs boson \cite{Aad:2012tfa,Chatrchyan:2012xdj}, the Standard Model (SM) is now complete. So far, it has proven to be extremely successful in describing particle physics data. However, the SM 
fails to explain several fundamental properties of the universe, like dark matter and the observed matter-antimatter asymmetry. It hence cannot be a complete theory, but has to be augmented by new physics (NP).

While there is no solid prediction for the mass scale of the new particles, the sensitivity of the electroweak scale to quantum corrections strongly suggests it to be not much beyond the TeV scale. Several promising models have been proposed in the past years, and the new particles are extensively searched for at the LHC. These searches however have not yet been successful, and the constraints on the masses of the new particles become more and more severe.

Direct searches for new particles in high energy collisions are however not the only opportunity to discover NP. Indirect methods, which probe high scales through their quantum corrections to low energy observables, offer alternative probes. Precision tests of the SM have the advantage that their NP reach is not limited by the energy of the experiment, therefore scales much beyond the TeV range can be probed. In order to achieve this, not only high experimental accuracy is required, but also very precise SM predictions for the observables in question.

A particularly promising arena for indirect NP searches is flavour physics. Flavour changing neutral current (FCNC) transitions are strongly suppressed in the SM, so that contributions from much higher scales have a chance to compete. The study of FCNC processes in the quark sector, however, is complicated by the fact that, due to QCD confinement, only rare decays of hadrons are observable in nature. A crucial ingredient for their precise theoretical description is therefore a good understanding of the involved hadronic effects. The latter, being goverened by low energy strong QCD interactions, are inaccessible to perturbation theory, and non-perturbative methods have to be used.

Among the most promising methods to tackle non-perturbative hadronic effects is their numerical evaluation on a discrete space-time lattice, commonly called lattice QCD. Lattice QCD results for the hadronic matrix elements entering rare meson decays have been relevant for flavour physics for many years. With the development of new simulation algorithms and the availability of increasing computer power, the accuracy of lattice results has seen impressive improvements, and some important processes have become accessible to lattice QCD calculations only recently. The importance of lattice QCD results for the phenomenology of quark flavour violation  has therefore significantly increased, and it is highly unlikely to reach its zenith anytime soon.

In this article I review the impact of recent lattice QCD calculations on the phenomenology of flavour and CP violating meson decays. 
In section \ref{sec:sm} I briefly recapitulate the basic ingredients of flavour and CP violation in the SM. Then in section \ref{sec:ckm} I review the status of the determination of the CKM mixing matrix from tree-level decays. 
Section \ref{sec:np} is devoted to some recent highlights of flavour physics to which lattice QCD has made substantial contributions. First I discuss the   recent lattice QCD results for the hadronic parameters entering neutral $B_{d,s}$ meson mixing and their implications for the flavour structure of NP contributions to meson mixing observables. Then I turn to the recent resurrection of direct CP violation in $K\to\pi\pi$ decays thanks to a first lattice QCD result for the relevant hadronic matrix elements, leading to a SM value of $\epe$ that is significantly below the data. Last but not least, I recapitulate the anomalies observed by the LHCb collaboration in semileptonic $b\to s$ transitions, where lattice QCD contributes to the determination of the $B\to K^{(*)}$ form factors. In section \ref{sec:sum} I give a short summary and outlook on future interesting directions for lattice QCD calculations enterning flavour phenomenology.

\section{Flavour physics in the SM}\label{sec:sm}

Flavour and CP violation in the SM is exclusively governed by the misalignment between the up and down Yukawa coupling matrices, described by the $3\times 3$ CKM matrix \cite{Cabibbo:1963yz,Kobayashi:1973fv}
\begin{equation}
V_\text{CKM} = \begin{pmatrix}
V_{ud} & V_{us} & V_{ub} \\
V_{cd} & V_{cs} & V_{cb} \\
V_{td} & V_{ts} & V_{tb} 
\end{pmatrix}\,.
\end{equation}
While due to the unitarity of $V_\text{CKM}$ neutral current interactions mediated by gluons, the photon and the $Z$ boson, remain flavour conserving at the tree-level, charged current interactions induced by the $W^\pm$ boson couplings are parametrised by the relevant elements of the CKM matrix and therefore induce flavour violating transitions. 

Flavour changing neutral current (FCNC) transitions in the SM are generated only through loop processes, by means of virtual $W^\pm$ boson and -- in the case of $K$ and $B$ meson decays -- up-type quark exchanges. Also in this case the unitarity of the CKM matrix leads to an effective suppression of FCNC transitions, due to the cancellation of all contributions that are independent of the quark masses running in the loop ({\it GIM mechanism}) \cite{Glashow:1970gm}.

\begin{figure}
\center{\includegraphics[width=.5\textwidth]{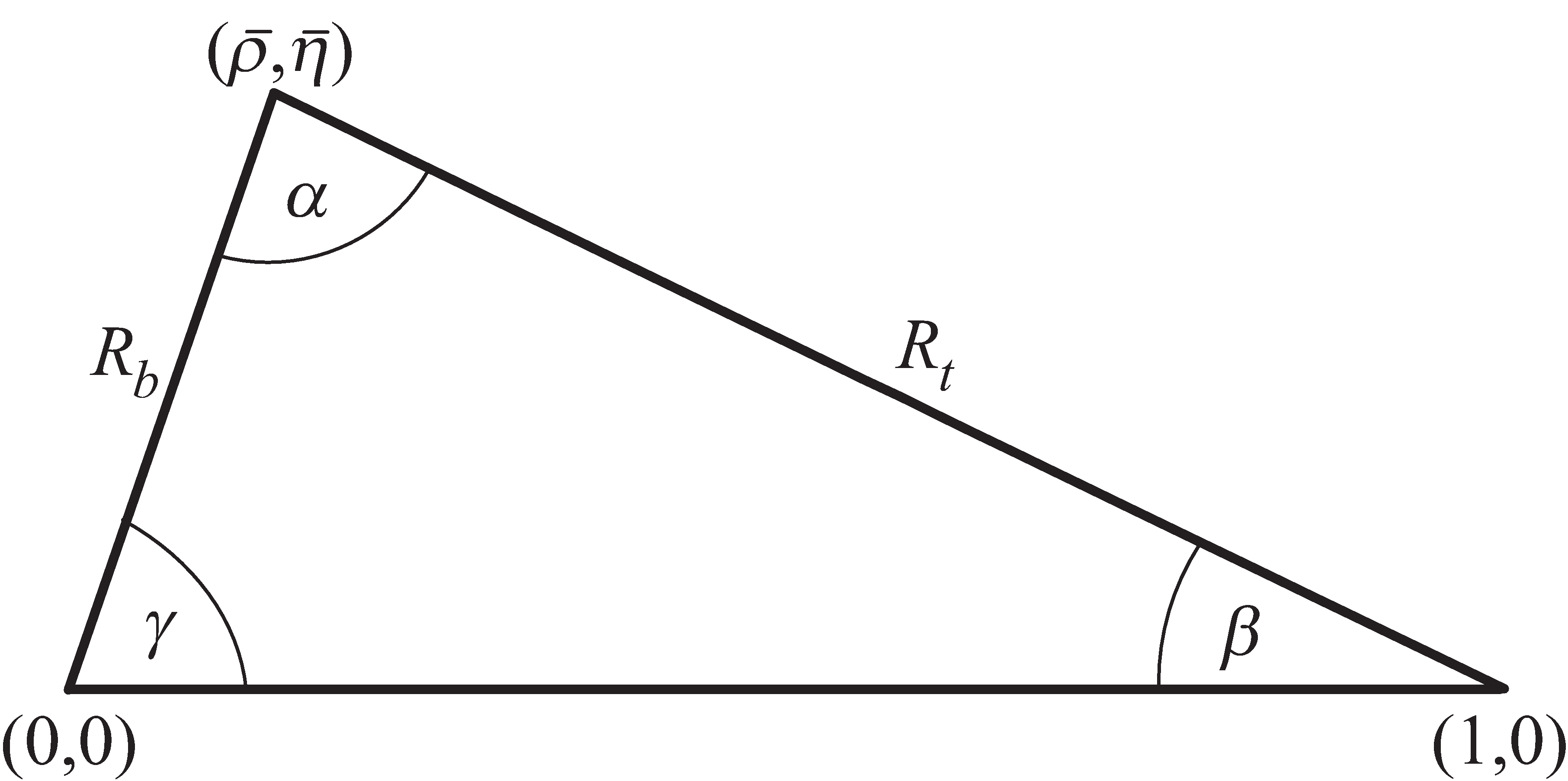}}
\caption{\it Unitarity triangle.}\label{fig:UT}
\end{figure}

The unitarity of the CKM matrix leads to several relations between its elements, among which
\begin{equation}
V_{ud}^{}V_{ub}^*+V_{cd}^{}V_{cb}^*+V_{td}^{}V_{tb}^*=0
\end{equation}
is the most popular one. Since all terms on the left hand side are of similar size, it can be conveniently displayed by a triangle in the complex plane, the so-called {\it unitarity triangle} shown in figure \ref{fig:UT}. With the base of the triangle normalised to unity, the sides $R_b$ and $R_t$ are given by
\begin{eqnarray}
R_b&=&\left|\frac{V_{ud}^{}V_{ub}^*}{V_{cd}^{}V_{cb}^*}\right|
= \left(1-\frac{\lambda^2}{2}\right) \frac{1}{\lambda}\frac{|V_{ub}|}{|V_{cb}|}\,,\\
R_t&=&\left|\frac{V_{td}^{}V_{tb}^*}{V_{cd}^{}V_{cb}^*}\right|=\frac{1}{\lambda}\frac{|V_{td}|}{|V_{cb}|}\,,
\end{eqnarray}
where $\lambda = |V_{us}|$ is the Cabibbo angle.

\section{CKM elements from tree-level decays}\label{sec:ckm}

The CKM matrix can be fully determined by the measurement of four independent parameters. A convenient determination is from tree-level charged current decays only, which can be used to find the mixing angles
\begin{equation}
|V_{us}|\,,\qquad |V_{cb}|\,, \qquad |V_{ub}|
\end{equation}
and the CP violating angle $\gamma$ of the unitarity triangle.  The advantage of determining the CKM matrix from tree-level decays lies in their insensitivity to NP contributions.
The obtained measurements therefore provide a model-independent determination of the CKM matrix, which is necessary in order to disentangle potential NP contributions to FCNC observables. In particular a precise knowledge of the element $|V_{cb}|$ is a crucial input for the SM predictions of several flavour precision observables like the CP violating parameter $\varepsilon_K$ in neutral kaon mixing, and the branching ratios of the very rare decays  $K\to\pi\nu\bar\nu$ and $B_s\to\mu^+\mu^-$.

The Cabibbo angle 
\begin{equation}
|V_{us}|= 0.2248\pm 0.0006
\end{equation}
 is precisely known from kaon decays \cite{PDG}, and the angle $\gamma$ can be determined from the CP asymmetry in $B\to DK$ decays with practically no theoretical uncertainty \cite{Brod:2013sga}. The accuracy of the current experimental determination by the LHCb collaboration \cite{LHCb-gamma}
\begin{equation}\label{eq:gamma}
\gamma = (72.2^{+6.8}_{-7.2})^\circ
\end{equation}
is limited mainly by statistics and will improve to $\sim 1^\circ$ precision with future LHCb and Belle II data. 

The determinations of $|V_{cb}|$ and $|V_{ub}|$ from semileptonic $B$ meson decays are theoretically more involved. Both mixing angles can be determined from inclusive decays, namely $B\to X_c\ell\nu$ and $B\to X_u\ell\nu$, respectively, with the results \cite{PDG,Gambino:2016jkc}
\begin{equation}\label{eq:incl}
|V_{cb}|_\text{incl} = (42.00 \pm 0.65) \cdot 10^{-3}\,,\qquad
|V_{ub}|_\text{incl} = (4.41 \pm 0.15^{+0.15}_{-0.19})\cdot 10^{-3}\,.
\end{equation}

The element $|V_{cb}|$ can also be determined from the exclusive decays $B\to D\ell\nu$ and $B\to D^*\ell\nu$. The current most precise exclusive value of  $|V_{cb}|$,
\begin{equation}
|V_{cb}|_{D^*\ell\nu} = (39.04\pm 0.75)\cdot 10^{-3}\,,
\end{equation}
is obtained from the latter decay, using a recent $N_f=2+1$ lattice calculation of the $B\to D^*$ form factor at zero recoil \cite{Bailey:2014tva}. Unfortunately at the moment only one unquenched lattice calculation is available for the form factor in question. An independent cross-check of the above result is badly needed, in particular in light of the apparent discrepancy with the inclusive determination of $|V_{cb}|$. 

For the $B\to D$ form factor at small recoil, on the other hand, results from several lattice collaborations using $N_f=2+1$ flavours are available \cite{Lattice:2015rga,Na:2015kha}. These can be extrapolated to the large recoil region, where more precise experimental data are available, by using
dispersion relations, unitarity constraints, and input from heavy quark effective theory. A fit to the available experimental and lattice results for $B\to D\ell\nu$ yields \cite{Bigi:2016mdz}
\begin{equation}
|V_{cb}|_{D\ell\nu} =  (40.49\pm0.97)\cdot 10^{-3}\,.
\end{equation}
Interestingly, this value lies in between the one determined from $B\to D^*\ell\nu$ and the inclusive measurement. 

The most precise determination of $|V_{ub}|$ is obtained from the exclusive decay $B\to\pi\ell\nu$. A recent $N_f=2+1$ lattice QCD calculation of the $B\to\pi$ form factors, combined with the model-independent $z$-expansion to the full $q^2$ range, yields the result \cite{Lattice:2015tia} (see also \cite{Dalgic:2006dt,Flynn:2015mha})
\begin{equation}
|V_{ub}|_{\pi\ell\nu} = (3.72\pm 0.16)\cdot 10^{-3}\,.
\end{equation}
This result is significantly below the inclusive value in \eqref{eq:incl}.
Note also that the PDG \cite{PDG} quotes an even lowevr value: $|V_{ub}|_{\pi\ell\nu} = (3.28\pm0.29)\cdot 10^{-3}$.

Last but not least, $|V_{ub}|$ and $|V_{cb}|$ can also be determined from the baryonic decays $\Lambda_b\to p\mu\nu$ and $\Lambda_b\to\Lambda_c\mu\nu$. The relevant form factors have been calculated on the lattice with $N_f = 2 + 1$ dynamical domain-wall fermion flavours \cite{Detmold:2015aaa}. Systematic uncertainties can be reduced by measuring the ratio of both decay rates. LHCb obtained \cite{Aaij:2015bfa}
\begin{equation}
\left|\frac{V_{ub}}{V_{cb}}\right|_{\Lambda_b\to p/\Lambda_c\mu\nu} = 0.083\pm0.04\pm0.04\,.
\end{equation}

\begin{figure}
\begin{center}
\includegraphics[width=.54\textwidth]{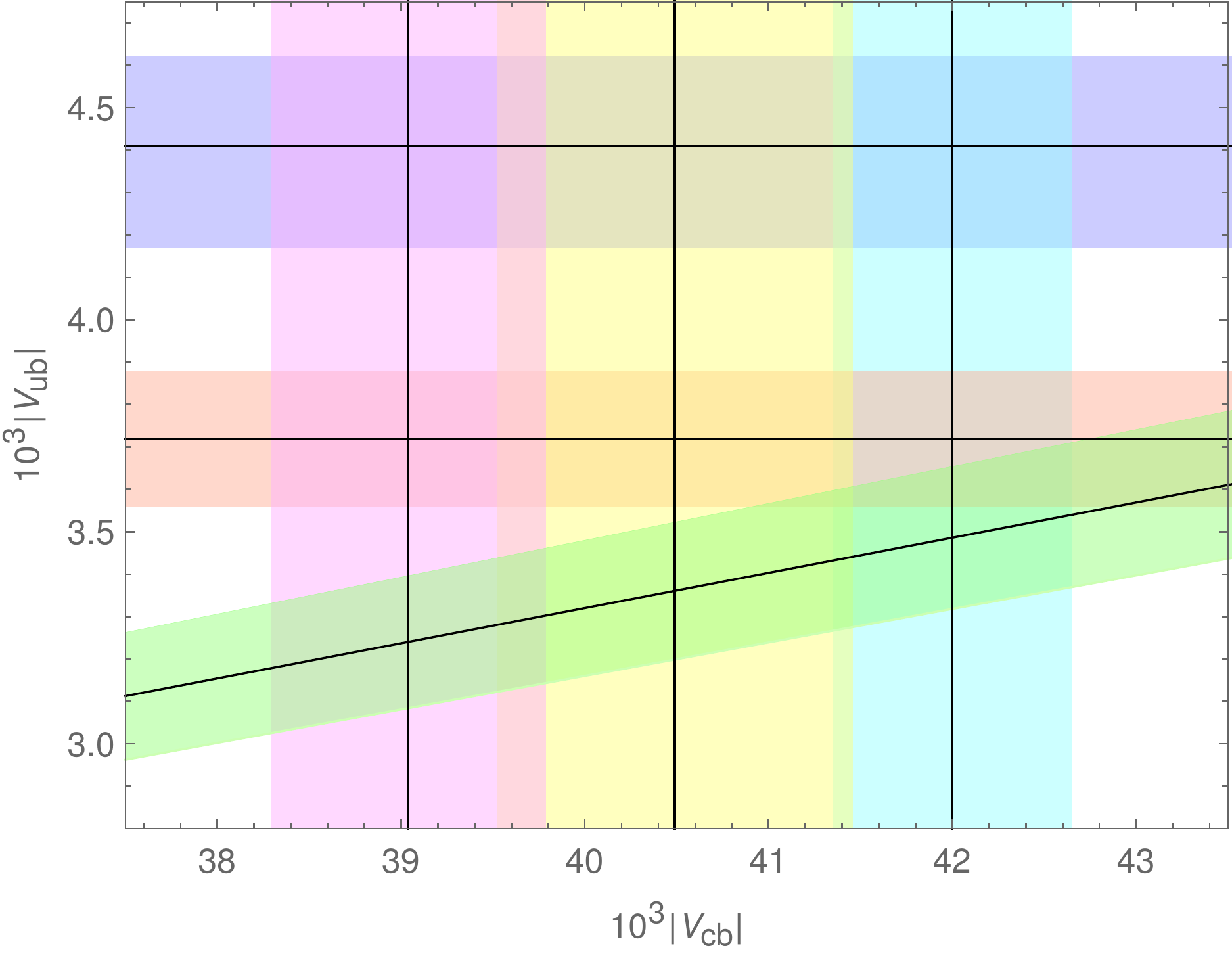}
\qquad
\includegraphics[width=.15\textwidth]{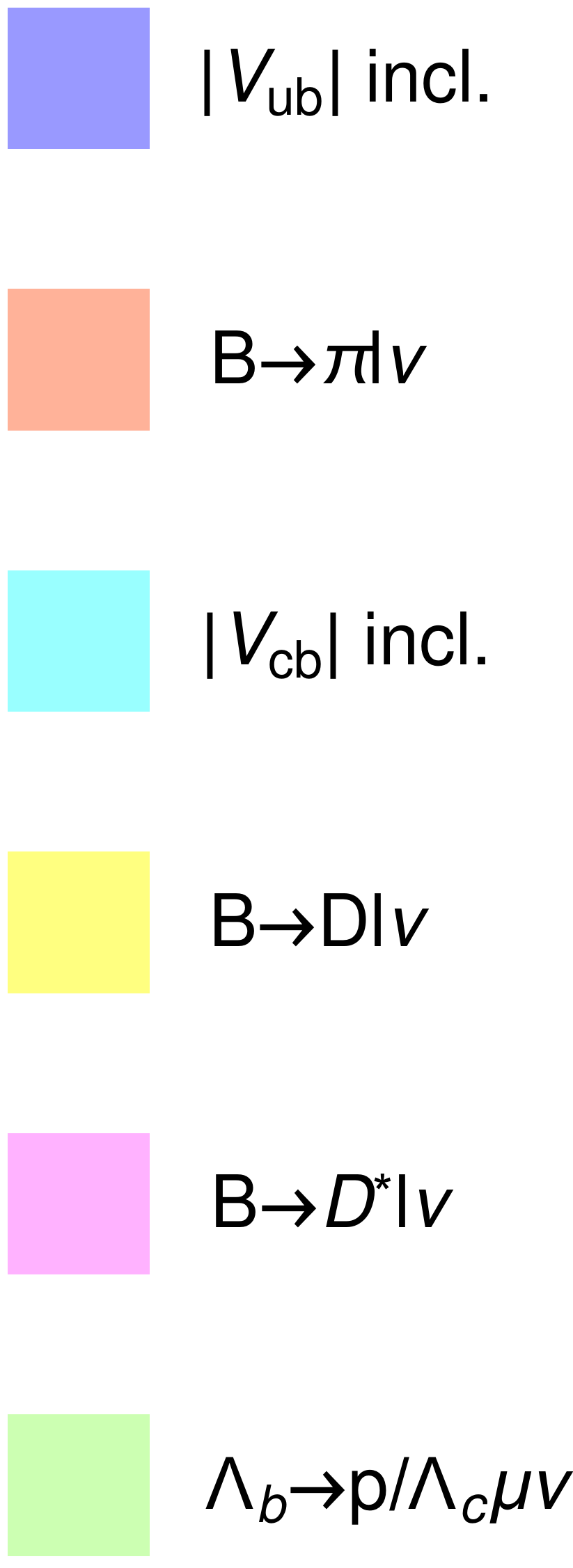}
\end{center}
\caption{Status of the determination of the CKM elements $|V_{ub}|$ and $|V_{cb}|$ from tree-level decays.\label{fig:vxb}}
\end{figure}

The present status of $|V_{ub}|$ and $|V_{cb}|$ determinations is summarised in figure \ref{fig:vxb}. As one can see, the situation is quite unsatisfactory at present, with significant discrepancies in both the determinations of $|V_{cb}|$ and $|V_{ub}|$. A major effort is therefore required on both the experimental and the theory side in order to improve the various determinations and to figure out what are the true values of these CKM elements. 

As mentioned above, the determination of CKM elements from tree-level decays is motivated by the assumed absence of significant NP contributions. With the current situation however, the question arises whether the origin of the observed tensions could actually be a non-negligible NP contribution. For example, the presence of right-handed charged currents would lead to non-universal results for $|V_{cb}|$ and $|V_{ub}|$ determinations \cite{Crivellin:2009sd,Buras:2010pz,Blanke:2011ry}. The observed deviations would follow the pattern
\begin{align}
& |V_{cb}|^{D\ell\nu} = |V_{cb}|(1+\delta)& \quad& |V_{ub}|^{\pi\ell\nu} = |V_{ub}|(1-\epsilon) \\
& |V_{cb}|^{D^*\ell\nu} = |V_{cb}|(1-\delta) && |V_{ub}|^{B\to\tau\nu} = |V_{ub}|(1+\epsilon) \\
& |V_{cb}|^\text{incl} = |V_{cb}|+\mathcal{O}(\delta^2) && |V_{ub}|^\text{incl} = |V_{ub}|+\mathcal{O}(\epsilon^2)
\end{align}
where $\delta$ and $\epsilon$ parametrise the right-handed contributions which enter the various decay modes in a non-universal manner. Unfortunately however, this pattern is not consistent with the observed deviations, so that right-handed currents can be excluded as the origin of the tension. Indeed a more general analysis including the effects from all possible dimension six operators lead to the conclusion that NP cannot explain the observed discrepancies in $|V_{cb}|$ and $|V_{ub}|$ determinations \cite{Crivellin:2014zpa}. On the other hand, a recent analysis of various $|V_{cb}|$ measurements considering muon and electron channels separately showed that contributions from the tensor operator may indeed be the origin of the tension in that case \cite{Colangelo:2016ymy}.

\section{New physics in flavour observables}\label{sec:np}

While the tree-level determination of CKM elements is expected to be insensitive to NP, significant NP contributions can arise in FCNC observables. Thanks to their strong suppression within the SM, they are sensitive to scales well beyond the energies of any present or planned high energy collider experiment. 

In order to gain the most insight from precision measurements of rare meson decay observables, a precise understanding of the SM contribution is mandatory. This requires not only accurate values for the CKM elements, as discussed in the previous section, but also a good control over the perturbative and non-perturbative contributions to the decay in question. For the latter, substantial progress has been made in recent lattice QCD calculations.

In what follows, some highlights of recent developments in flavour physics are presented, which all exhibit some tension between the data and the corresponding SM predictions. It is common to all cases that lattice QCD plays a substantial role, and future improved results will help to clarify the situation.

\subsection{$B_{s,d}-\bar B_{s,d}$ mixing}

Earlier this year, the Fermilab Lattice and MILC collaborations presented a new determination of the hadronic matrix elements entering $B_d-\bar B_d$ and $B_s-\bar B_s$ mixing \cite{Bazavov:2016nty}. Their results
\begin{eqnarray}
f_{B_d}^2{\hat B_{B_d}} &=& (0.0518\pm0.004\pm0.0010)\gev^2\,, \\
 f_{B_s}^2{\hat B_{B_s}} &=& (0.0754\pm0.0046\pm0.0015)\gev^2 \,,\\
\xi &=& \frac{f_{B_s}\sqrt{\hat B_{B_s}}}{f_{B_d}\sqrt{\hat B_{B_d}}} = 1.206\pm0.018\pm0.006
\end{eqnarray}
not only are more precise than previous lattice determinations \cite{Aoki:2016frl}, but it also turned out that with the new input the SM predictions for the mass differences $\Delta M_d$ and  $\Delta M_s$ differ from the experimental values by $1.8\sigma$ and $1.1\sigma$, respectively. The ratio 
$\Delta M_d/\Delta M_s$, due to the smaller uncertainty in the parameter $\xi$, is even  $2.0\sigma$ below the data.

At this stage, it is certainly too early to claim the presence of NP in $B_{s,d}-\bar B_{s,d}$ mixing, and an independent cross-check of the above results by a different lattice collaboration would be more than welcome. However it is instructive to investigate what this tension, if eventually confirmed, would imply for the flavour structure of the NP entering $B_{s,d}-\bar B_{s,d}$ mixing observables. 

The simplest extension of the SM, as far as flavour physics is concerned, are models with Constrained Minimal Flavour Violation (CMFV) \cite{Buras:2000dm,Buras:2003jf,Blanke:2006ig}. In this setup, it is assumed that flavour and CP symmetries are, as in the SM, only violated by the Yukawa couplings $Y_d$ and $Y_d$. Consequently the CKM matrix parametrises all flavour and CP violating effects also in the NP sector. In addition the effective operator structure describing flavour violating transitions is taken to be the same as in the SM.\footnote{This latter assumption distinguishes CMFV models from the more general concept of Minimal Flavour Violation derived from an effective field theory ansatz \cite{D'Ambrosio:2002ex}.}

CMFV is a phenomenologically appealing ansatz, because it provides an effective suppression of TeV-scale NP contributions in FCNC observables. Furthermore, it is a very predictive framework. Since all flavour violating interactions are proportional to the same CKM elements as in the SM, the relative size of NP effects is flavour universal. Therefore CMFV contributions to FCNC observables can be parametrised simply by replacing the SM loop functions by new real and flavour universal functions. For instance, in the case of meson mixing ($\Delta F = 2$), $S_0(x_t)$ is replaced by
\begin{equation}
S(v) = S_0(x_t) + \Delta S(v) \ge S_0(x_t) = 2.322\,,
\end{equation}
where $v$ collectively denotes the parameters of the NP sector. The NP contributions to $\Delta F =2 $ observables in CMFV models can thus be described by a single real and flavour universal parameter $\Delta S(v)$.
The lower bound $\Delta S(v)\ge 0$ is a consequence of the $\Delta F =2$ amplitude structure in CMFV models~\cite{Blanke:2006yh}.

\begin{figure}
\includegraphics[width = 0.47\textwidth]{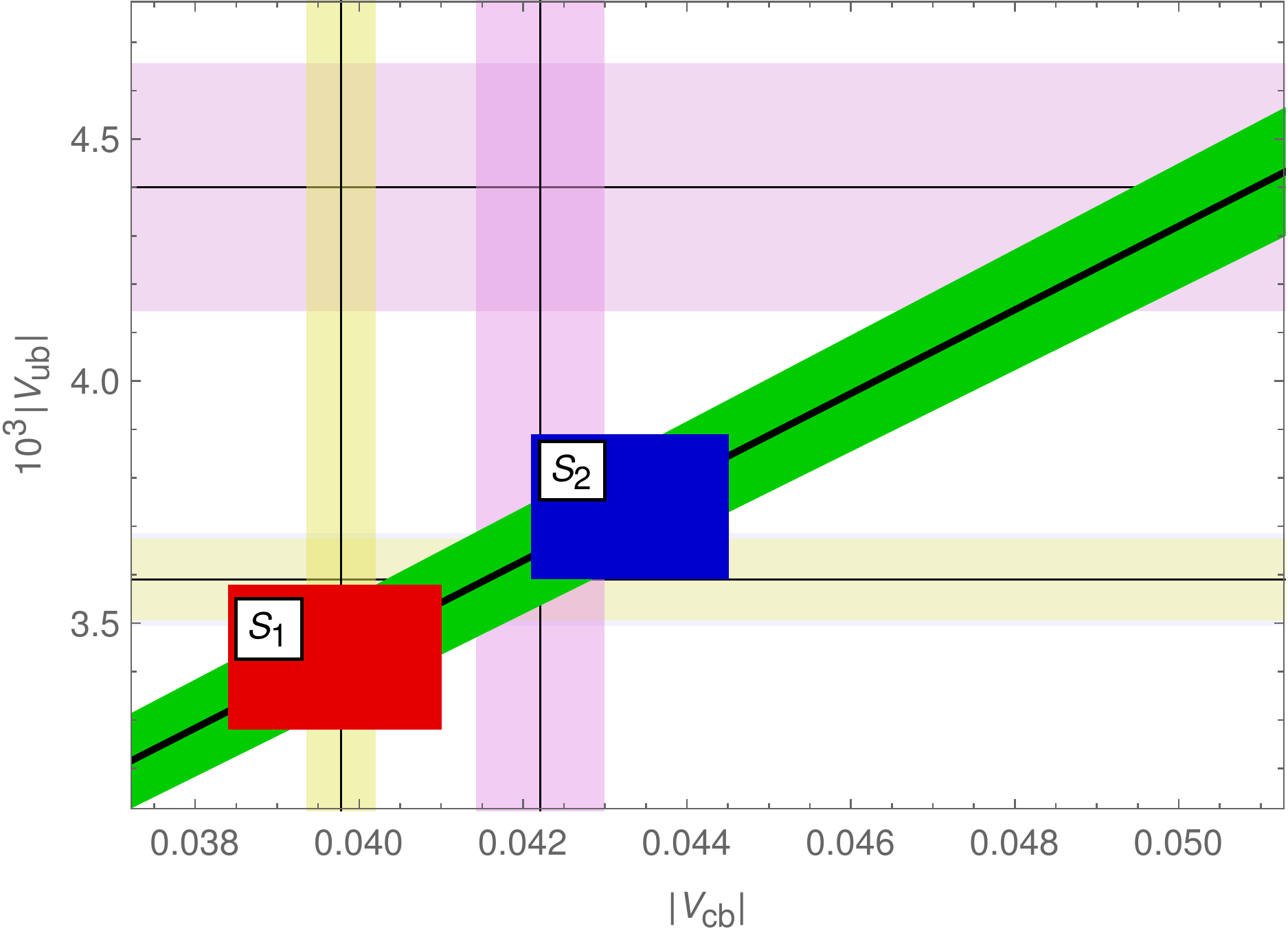}\hfill
\includegraphics[width = 0.48\textwidth]{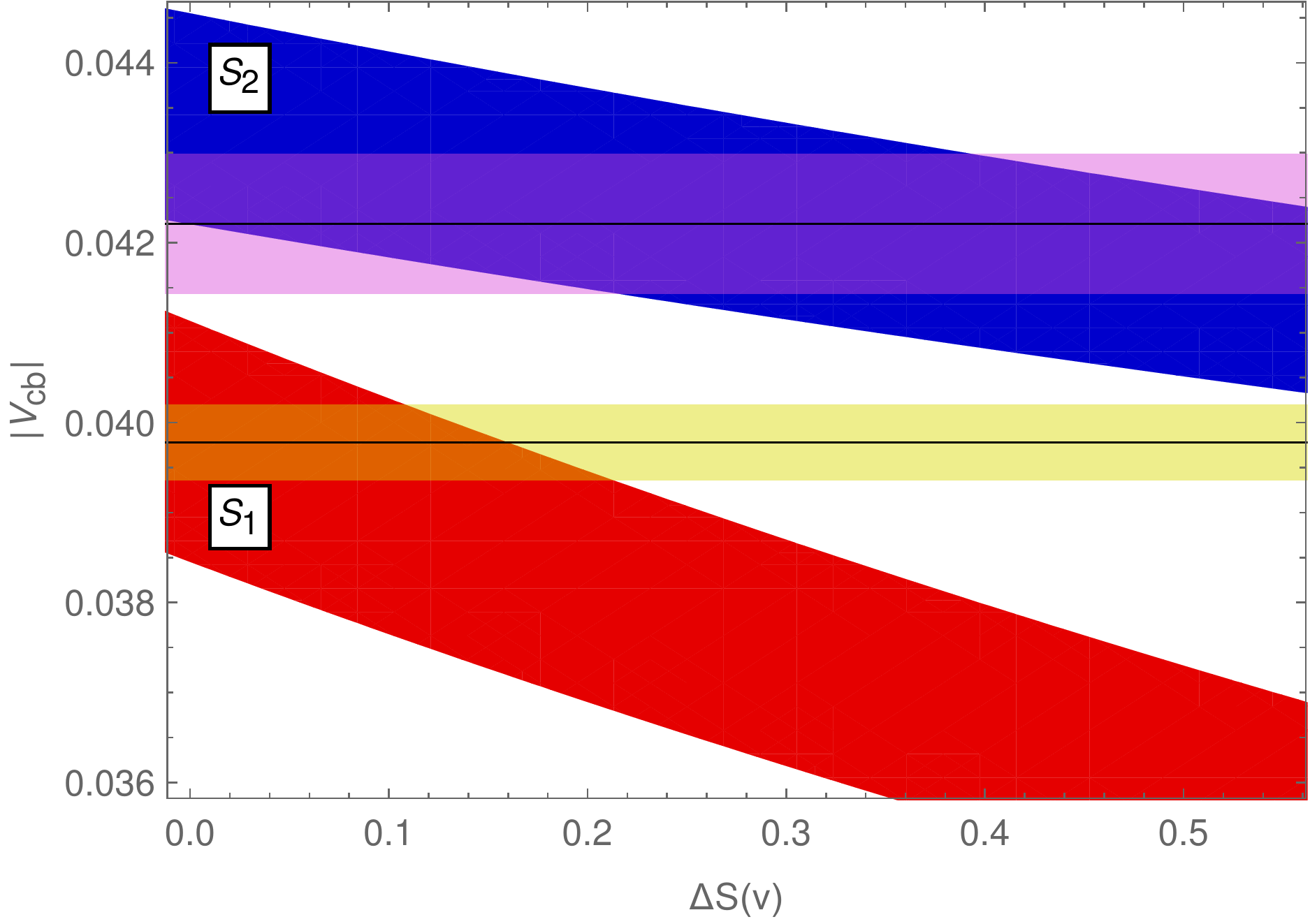}
 \caption{{\it Left:} $\vub$ versus $\vcb$ in CMFV (green) compared with the tree-level 
exclusive (yellow) and inclusive (violet) determinations summarised in section \protect\ref{sec:ckm}. The squares display the results in $S_1$ (red) and $S_2$ (blue). {\it Right:}  $\vcb$ versus {the flavour-universal NP contribution} $\Delta S(v)$ obtained in $S_1$ (red) and $S_2$ (blue). The horizontal bands correspond to the tree-level measurements of $\vcb$. Figures taken from \protect\cite{Blanke:2016bhf}.\label{fig:Bs}}
\end{figure}

In CMFV models, the unitarity triangle can not only be determined from tree-level decays, but also from FCNC observables that remain unaffected by CMFV. These are the time-dependent CP asymmetry $S_{\psi K_S}$ that measures the CP violating phase $2\beta$ of $B_d-\bar B_d$ mixing, and the ratio $\Delta M_d/\Delta M_s$ that determines the length of the side $R_t$. Together with $|V_{us}|$ from tree-level decays, these measurements leave us with the {\it universal unitarity triangle (UUT)} that holds within all CMFV models \cite{Buras:2000dm}. Using the most recent lattice and other input parameters, one finds for the apex of the UUT \cite{Blanke:2016bhf}
\begin{equation}
\bar\rho_\text{UUT} = 0.170\pm 0.013\,,\qquad \bar\eta_\text{UUT} = 0.333 \pm 0.011\,,
\end{equation}
reaching a few percent precision. Consecutively, the determination of the angle $\gamma$ yields
\begin{equation}\label{eq:gammaCMFV}
\gamma_\text{UUT} = (63.0 \pm 2.1)^\circ\,,
\end{equation}
to be compared with the tree-level measurement in \eqref{eq:gamma}. From the length of the side $R_b$ one obtains
\begin{equation}
\left|\frac{V_{ub}}{V_{cb}}\right|_\text{UUT} = 0.0864 \pm 0.0025\,.
\end{equation}
As shown in the left panel of figure \ref{fig:Bs}, this relation clearly disfavours the inclusive value for $\vub$. 

In order to fully determine the CKM matrix, one additional input is needed to fix the value of $\vcb$ and to, in turn, make predictions for other observables. Within the $\Delta F=2$ sector there are two quantities that are well suited, as they are precisely known and strongly dependent on the value of $\vcb$. These are the mass difference $\Delta M_s$ in the $B_s$ system and the parameter $\eps_K$ that describes CP violation in $K^0-\bar K^0$ mixing. This defines the two strategies:\vspace{2mm}

{\bf\boldmath Strategy $S_1$:} The experimental value of 
$\Delta M_s$ is used to determine $\vcb$ as a function of $S(v)$.\vspace{2mm}

{\bf\boldmath Strategy $S_2$:}
The experimental value of 
$\varepsilon_K$ is used to fix $\vcb$ as a function of $S(v)$.\vspace{2mm}

The result of this determination is shown in the right panel of figure \ref{fig:Bs}. showing $\vcb$ as a function of the NP contribution $\Delta S(v)>0$. It is apparent that the determinations of $\vcb$ from $\Delta M_s$ and $\eps_K$ are inconsistent with each other, and that the tension is smallest within the SM, i.\,e.\ when $\Delta S(v)=0$ \cite{Blanke:2016bhf}.

It then needs to be addressed how the tension in $\Delta F=2$ data, if eventually confirmed by future lattice results, could be resolved by NP. As the CMFV framework, i.\,e.\ a flavour universal positive NP contribution to $\Delta F =2$ observables, is not successful, two routes are in principle possible. While difficult to achieve in concrete models, it is in principle possible to relax the lower bound $\Delta S(v)\ge0$ while keeping the function $S(v)$ flavour universal. It is then possible to find a consistent solution for $\vcb$ from $\Delta M_s$ and $\eps_K$, however the obtained value is then inconsistent with the tree-level determination of that CKM parameter. 

Abandoning the CMFV hypothesis and allowing for the presence of new sources of flavour and CP violation,  the flavour universality in $\Delta F=2$ observables is violated. In this case the NP contributions can be parametrised by replacing the function $S_0(x_t)$ by the flavour dependent complex functions
\begin{equation}
S_i=|S_i| e^{i\varphi_i}, \qquad  i=K,s,d \,.
\end{equation}
With six independent new parameters it is always possible to fit the available $\Delta F =2$ data, in agreement with the tree-level determination of the CKM matrix. In concrete NP models, the correlations between $\Delta F =2$ and $\Delta F =1$ observables can then be used to test  a given model.

More predictive already within the $\Delta F=2$ sectot are models with a minimally broken $U(2)^3$ flavour symmetry \cite{Barbieri:2011ci,Barbieri:2012uh,Buras:2012sd,Blanke:2016bhf}. In that case, $K^0-\bar K^0$ mixing can only be enhanced with respect to the SM, and the contributions to $B_d-\bar B_d$ and $B_s-\bar B_s$ mixing are flavour universal:
\begin{equation}
S_K=r_K S_0(x_t)\ \text{ with } r_K\ge 1\,,\qquad
 S_d=S_s=r_B S_0(x_t) e^{i\varphi_{B}}\,.
\end{equation}
With this structure it is possible to fit the currently available $\Delta F =2 $ data. The scenario can however be tested with tree-level and $\Delta F =2$ data alone, by using the triple correlation between the CP asymmetry $S_{\psi K_S}$, the phase $\phi_s$ of $B_s-\bar B_s$ mixing and the ratio $|V_{ub}/V_{cb}|$ \cite{Buras:2012sd}. Also the low value of $\gamma$ in \eqref{eq:gammaCMFV}, that also holds in $U(2)^3$ models, may turn out to be problematic one day.

\subsection{Direct CP violation in kaon decays}

The description of direct CP violation in $K\to\pi\pi$ decays, quantified by the parameter $\epe$, has been a long-standing open issue in the SM. While a precise experimental value \cite{Batley:2002gn,AlaviHarati:2002ye,Abouzaid:2010ny},
\begin{equation}
\text{Re}(\epe)_\text{exp} = (16.6\pm 2.3)\cdot 10^{-4}\,,
\end{equation}
existed since the early 2000s, until recently no solid SM prediction had been available. A major step forward has been made with the first lattice calculation of the hadronic matrix elements
\begin{equation}\label{eq:B6B8}
B_6^{(1/2)} = 0.57\pm 0.19\,, \qquad B_8^{(3/2)} = 0.76\pm 0.05
\end{equation}
by the RBC-UKQCD collaboration \cite{Bai:2015nea}, hinting to a SM value for $\epe$ that is significantly lower than the measured value. 
While in the strict large $N_c$ limit, $B_6^{(1/2)}$ and $B_8^{(3/2)}$ are equal to unity, a recent analysis within the dual QCD approach resulted in the bound \cite{Buras:2015xba,Buras:2016fys}
\begin{equation}\label{eq:dual}
B_6^{(1/2)} < B_8^{(3/2)} < 1\,,
\end{equation}
supporting the RBC-UKQCD result \eqref{eq:B6B8}. It is interesting to note that if future more precise lattice calculations confirm the bound \eqref{eq:dual}, the presence of NP in $\epe$ will unambiguously be confirmed. This result triggered a number of phenomenological analyses, both in and beyond the SM.

Within the SM, a simple phenomenological expression can be derived \cite{Buras:2003zz,Buras:2015yba}
\begin{equation}
\text{Re}(\epe)=\frac{\text{Im}(V_{ts}^* V^{}_{td})}{1.4\cdot 10^{-4}}\cdot 10^{-4} \cdot\left[\left(
-3.6+21.4 \,B_6^{(1/2)} \right) + \left( 1.2-10.4 \, B_8^{(3/2)} \right) \right] \,.
\end{equation}
The first term in the square brackets represents the $\Delta I =1/2$ amplitude that is mainly generated by QCD penguin contributions. The $\Delta I =3/2$ contribution, mainly due to electroweak penguins, leads to the second term in the square brackets. The two contributions cancel each other to a large extent, leading to a very small SM prediction. 

Two independent analyses have evaluated the SM prediction at next-to-leading order, taking into account the recent lattice result \cite{Bai:2015nea}, with consistent results
\begin{equation}
\text{Re}(\epe)_\text{SM} = (1.9\pm4.5)\cdot 10^{-4} \quad\text{\cite{Buras:2015yba}}\,,\qquad
\text{Re}(\epe)_\text{SM} = (1.06\pm5.07)\cdot 10^{-4} \quad\text{\cite{Kitahara:2016nld}}\,.
\end{equation}
These results exhibit a $2.9\sigma$ tension with the data. A complete next-to-next-to-leading order calculation is currently in progress \cite{epe-NNLO}.

\begin{figure}
\centering{\includegraphics[width=.55\textwidth]{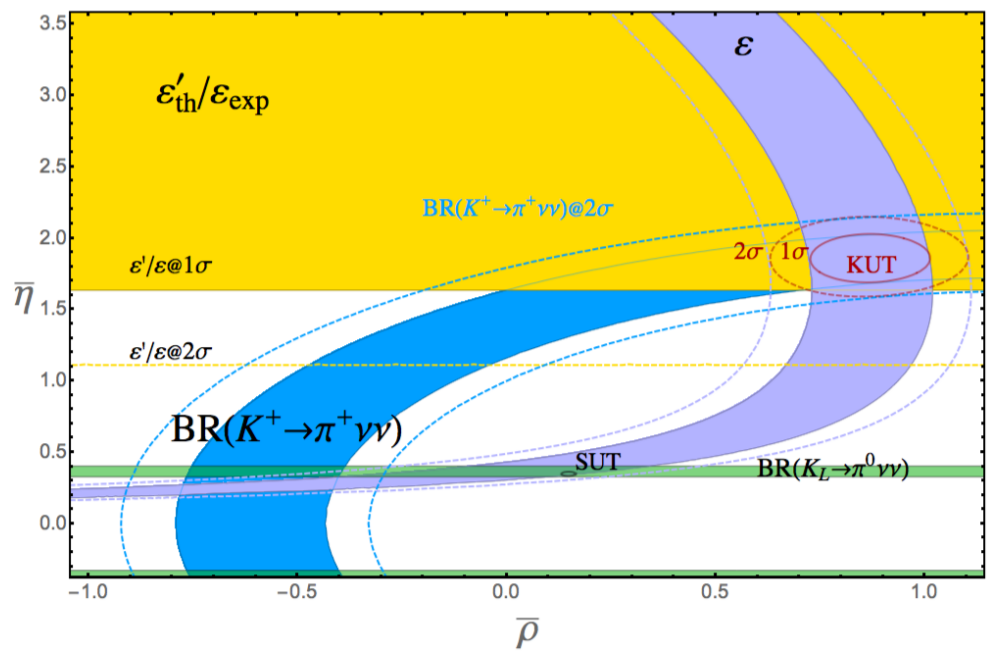}
}
\caption{$K$-unitarity triangle extracted from a scenario with present central values for $K$ decay observables (SM value for $\mathcal{B}(K_L\to\pi^0\nu\bar\nu)$) and uncertainties reduced to $\pm 10\%$. Figure taken from \protect\cite{Lehner:2015jga}. \label{fig:KUT}}
\end{figure}

Having at hand a number of kaon decay observables that are theoretically well understood and precisely measured, allows to construct the unitarity triangle from $K$ physics only ($K$-unitarity triangle) \cite{Buras:1994ec,Buchalla:1994tr,Lehner:2015jga}. Apart from $\eps_K$, which is already known to be an important player in global unitarity triangle fits \cite{Charles:2015gya,Bevan:2014cya}, important imput will come from the branching ratios of the extremely clean decays $K^+\to\pi^+\nu\bar\nu$ and $K_L\to\pi^0\nu\bar\nu$. Last but not least, with future improved lattice determinations of the hadronic matrix elements $B_6^{(1/2)}$ and $B_8^{(3/2)}$, $\epe$ will be crucial to overconstrain the $K$-unitarity triangle. A mismatch in the $K$-unitarity triangle constraints as indicated in figure \ref{fig:KUT} would be a clear signal of NP present in the kaon sector.

Due to the strong suppression of $\epe$ in the SM, NP effects can lead to large enhancements of this observable. Following the recent lattice results, $\epe$ has been analysed in a number of 
NP models \cite{Blanke:2015wba,Buras:2015yca,Buras:2015jaq,Buras:2015kwd,Buras:2016dxz,Tanimoto:2016yfy,Kitahara:2016otd,
Bobeth:2016llm}.
In \cite{Buras:2015yca}, the correlation between $\epe$, $\eps_K$ and $K\to\pi\nu\bar\nu$ has been analysed in the context of simplified models with tree-level flavour changing $Z$ or $Z'$ couplings. If the latter couplings are purely left-handed, strong correlations between the four observables in question can be found. The stringent constraint from the measured value of $\eps_K$ leads to a correlation between the rare decays $K^+\to\pi^+\nu\bar\nu$ and $K_L\to\pi^0\nu\bar\nu$, allowing for two branches in the $K\to\pi\nu\bar\nu$ plane, see also \cite{Buras:2012jb}.  This correlation is well knwon from other models with only left-handed flavour violating interactions, and has been shown model-independently to be a consequence of the effective operator structure in $\Delta S =2$ and $\Delta S=1$ transitions \cite{Blanke:2009pq}.
Both $K_L\to\pi^0\nu\bar\nu$ and $\epe$ are sensitive to the CP violating phase of the $Z$ or $Z'$ coupling, however in the case of purely left-handed interactions, the new contributions enter with a relative minus sign. Therefore the enhancement of $\epe$ needed to reconcile that observable with the data implies a suppression of $K_L\to\pi^0\nu\bar\nu$ and will make the latter decay very challenging to observe experimentally. In the case of both left- and right-handed flavour changing $Z^{(\prime)}$ couplings, however, a simultaneous enhancement of the two observables is possible. Also the correlation between the two $K\to\pi\nu\bar\nu$ decays is lost.

Models with an $S(3)_c\times SU(3)_L\times U(1)$ gauge symmetry provide concrete realisations of tree-level flavour changing $Z$ and $Z'$ couplings. An update of their predictions for FCNC observables has been presented in \cite{Buras:2015kwd,Buras:2016dxz} in light of the $\epe$ anomaly. In addition to the general conclusions from the simplified models discussed above, in these scenarios also predictions for rare $B$ decays can be derived, to be confronted with future data by LHCb, CMS and Belle II.

\begin{figure}
\includegraphics[width=.48\textwidth]{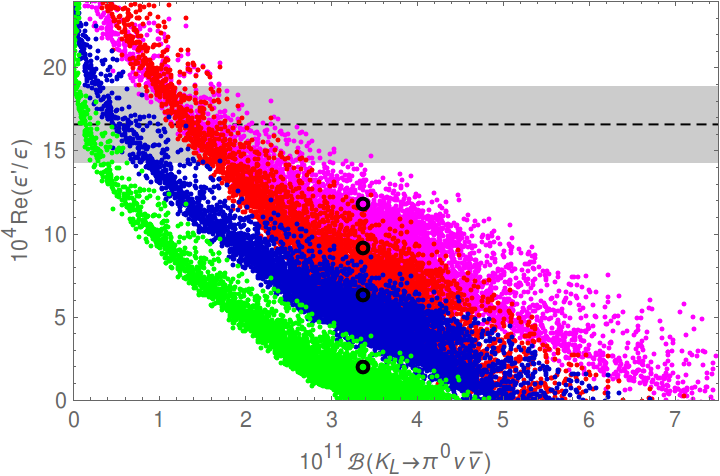}\hfill
\includegraphics[width=.48\textwidth]{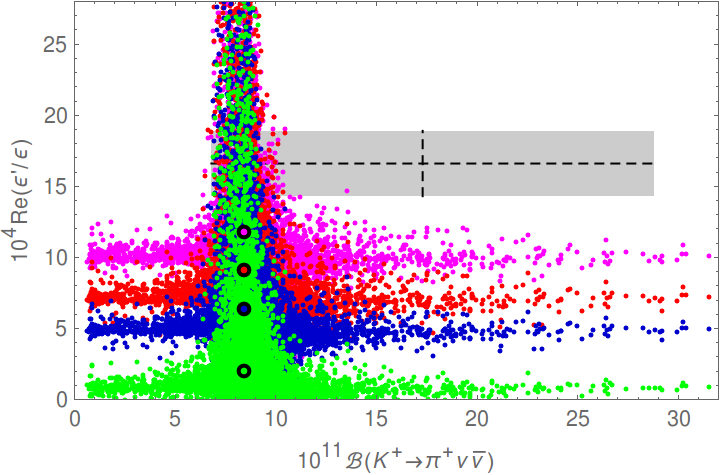}
\caption{\label{fig:LHT}Correlation between the $K\to\pi\nu\bar\nu$ branching ratios
 and $\text{Re}(\epe)$ in the LHT model, setting the NP scale $f$ to $1\tev$. The colours depict different choices for the hadronic matrix elements $(B_6^{(1/2)},B_8^{(3/2)})$: $(1.0,1.0)$ (red), $(0.76, 0.76)$ (blue), $(0.57,0.76)$ (green), $(1.0,0.76)$ (magenta). The black dots show the corresponding central SM values. Figures taken from \protect\cite{Blanke:2015wba}.}
\end{figure}

A prominent example of NP with only left-handed flavour violating interactions is the Littlest Higgs model with T-parity (LHT), whose flavour phenomenology has been analysed extensively over the past decade \cite{Hubisz:2005bd,Blanke:2006sb,Blanke:2006eb,Blanke:2007wr,Bigi:2009df,Blanke:2009am}. Recently an update has been presented \cite{Blanke:2015wba}, taking into account the constraints from direct new particle searches at the LHC as well as the experimental and theoretical improvements in the flavour sector. Figure \ref{fig:LHT} shows the correlation of $\epe$ with the rare decays $K_L\to\pi^0\nu\bar\nu$ and $K^+\to\pi^+\nu\bar\nu$, respectively. Again, due to the pure left-handedness of FCNC processes, an enhancement of $\epe$ governed by electroweak penguin contributions implies a suppression of $\mathcal{B}(K_L\to\pi^0\nu\bar\nu)$. Furthermore, due to the correlation between the charged and neutral $K\to\pi\nu\bar\nu$ decay rates, NP in $\epe$ implies a SM-like $K^+\to\pi^+\nu\bar\nu$ branching ratio. 

The case of only SM operators contributing to rare kaon decay observables has been analysed in a model-independent way in \cite{Buras:2015jaq}. A particularly interesting correlation emerges
if NP cures the $\epe$ anomaly through new contributions to the QCD penguins: In that scenario the correlation between $\klpn$ and $\kpn$ will be confined to a single branch parallel to the Grossman-Nir bound \cite{Grossman:1997sk}, simultaneously enhancing $\klpn$ and $\kpn$. Moreover, the NP contribution to $\Delta M_K$
will turn out to be negative. If, on the other hand, NP cures the $\epe$ anomaly through electroweak penguin contributions, $\Delta M_K$ will be enhanced. While at present, possible NP effects in $\Delta M_K$ having either sign are possible, due to the poorly known long-distance contributions in the SM, future lattice calculations of the latter will be decisive on this aspect.

Last but not least, also in the MSSM it is possible to reconcile $\epe$ with the data \cite{Tanimoto:2016yfy,Kitahara:2016otd}. Dangerously large contributions to indirect CP violation in kaon mixing, $\eps_K$, can in this case be avoided by choosing the mass ratio $m_{\tilde g}/M_S\gsim 1.5$, where $M_S = m_{\tilde Q} = m_{\tilde D} = \mu_\text{SUSY}$. The anomaly in $\epe$ can then be explained with squark masses in the multi-TeV regime, that are difficult to access directly in high-$p_T$ searches \cite{Kitahara:2016otd}. 

\subsection{$b\to s$ penguin transitions}

Another powerful  test of the SM flavour sector is given by observables measuring $b\to s$ penguin transitions. As $B$ mesons can only decay via weak interactions and the decay rate is suppressed by a factor of $|V_{cb}|^2$, FCNC decays appear with significant branching ratios and are experimentally relatively easy to access. Additionally, a large number of final states is accessible. Over the past years, an appealing pattern of tensions with the SM predictions has emerged in the data on semileptonic $b\to s\ell^+\ell^-$ transitions. 

Most striking in this context are a $3.4\sigma$ deviation of the angular observable $P'_5$ of the decay $B\to K^*\mu^+\mu^-$ measured by LHCb \cite{Aaij:2015oid} and confirmed, albeit with lower statistics, by Belle \cite{Abdesselam:2016llu}, as well as a $2.6\sigma$ tension in the ratio \cite{Aaij:2014ora}
\begin{equation}
R_K = \frac{\mathcal{B}(B\to K\mu^+\mu^-)}{\mathcal{B}(B\to Ke^+e^-)}\,,
\end{equation}
which provides a test of lepton flavour universality. Interestingly, a number of other data, like a recent Belle measurement of $P'_5$ in the $B\to K^*\mu^+\mu^-$ and $B\to K^*e^+e^-$ channels separately \cite{Wehle:2016ckm}, hint in the same direction, although the observed tension is not statistically significant. 

From the theory point of view, $b\to s\ell^+\ell^-$ transitions can be conveniently described by the effective Hamiltonian \cite{Buchalla:1995vs}
\begin{equation}\label{eq:Heff-bs}
\mathcal{H}_\text{eff}= -\frac{4 G_F}{\sqrt{2}} V_{tb}^* V_{ts}^{} \frac{e^2}{16\pi^2}\sum_i(C_i {\cal O}_i +C'_i {\cal O}'_i)+h.c.\,.
\end{equation}
In the SM, only the unprimed Wilson coefficients $C_i$, corresponding to left-handed FCNC transitions, are relevant, due to the left-handedness of the flavour violating weak interactions. NP contributions, on the other hand, can have either chirality. The operators most sensitive to NP are the dipole operators $\mathcal{O}^{(\prime)}_7$ and the four fermion operators $\mathcal{O}^{(\prime)}_9$ and $\mathcal{O}^{(\prime)}_{10}$ that are not affected by tree-level contributions in the SM.

Various collaborations \cite{Altmannshofer:2015sma,Capdevila:2016fhq,Beaujean:2015gba} have performed global fits of the Wilson coefficients in \eqref{eq:Heff-bs}, with the concordant conclusion that the data on $b\to s\mu^+\mu^-$ transitions exhibit a tension of $\sim 4\sigma$ with the SM, and that the tension can be explained with a NP contribution
\begin{equation}\label{eq:deltaC9}
\delta C_9^\text{NP} \simeq -1 \,.
\end{equation}
If this NP contribution is assumed to affect only the muon channel, but not the electron channel, then also the observed deviation in $R_K$ can be accounted for. 

Unfortunately, many observables in the $b\to s\ell^+\ell^-$ system are affected by non-perturbative uncertainties that are difficult to control in the SM. On the one hand, the form factors describing the $B\to K^*$ (or similar) transition are affected by long-distance QCD effects. At low $K^*$ recoil, these form factors can be accessed by lattice QCD, while at large recoil currently other non-perturbative methods, for example light-cone sum rules, are required. On both ends systematic improvements are possible, and more accurate theoretical predictions for the form factors will significantly reduce the theoretical uncertainties in observables like $P'_5$. 

An additional source of theoretical uncertainty are non-factorisable contributions that cannot be captured by the form factors. They are attributed to low energy charm- and up-quark loops that contribute mainly below the $c\bar c$ threshold. Typically these contributions will affect the Wilson coefficient $C_9$, thus making it difficult to disentangle the NP effect in \eqref{eq:deltaC9} from a possible non-factorisable SM contribution that has so far been underestimated. Unfortunately for the time being, no solid theory prediction is available for the size of these sontributions. It remains to be seen whether one day it will be possible to evaluate these non-factorisable effects on the lattice. 

For the time being, the only way to disentangle a possible NP effect in $C_9$ from non-factorisable SM effects are more precise measurements. On the one hand, if  a lepton flavour non-universal effect is confirmed in $R_K$ and in further measurements comparing $B\to K^*\mu^+\mu^-$ with $B\to K^*e^+e^-$, it will be clear that QCD cannot account for the observed tension. On the other hand, if future global fits show that the data cannot be explained by a $q^2$-independent new contribution\footnote{Here $q^2$ denotes the invariant mass squared of the final state lepton pair, so that a low $q^2$ corresponds to a large recoil of the final state meson and vice versa.} to the Wilson coeffcient~$C_9$, then the interpretation in terms of non-factorisable effects will be favoured.

Last but not least, an important player in the $b\to s\ell^+\ell^-$ system is the rare decay $B_s\to\mu^+\mu^-$. The latter decay has been discovered by a joint effort of the LHCb and CMS collaborations, with the result \cite{CMS:2014xfa}
\begin{equation}
\overline{\mathcal{B}}(B_s\to\mu^+\mu^-) = \left(2.8^{+0.7}_{-0.6}\right) \cdot 10^{-9}\,.
\end{equation}
This measurement, while a bit on the low side, is consistent with the SM prediction at the $2\sigma$ level. Significant improvements in the experimental accuracy can be expected over the coming years, as the uncertainty is currently dominated by the low statistics. 

Theoretically, the $B_s\to\mu^+\mu^-$ decay is rather clean. In contrast to the semileptonic decays discussed above, this mode is not affected by form factor uncertainties, but all non-perturbative effects can be summarised in the decay constant $f_{B_s}$, for which precise lattice calculations are available \cite{Aoki:2016frl} and further improvements can be expected. The main uncertainty currently lies in the determination of the CKM element $|V_{cb}|$, as the $B_s\to\mu^+\mu^-$ branching ratio is very sensitive to that value. Further improvements of the accuracy of $|V_{cb}|$ are therefore badly needed also from the point of view of rare $B$ decays.

\section{Summary and outlook}\label{sec:sum}

Flavour physics has entered the precision era, and its story is far from being complete. Further efforts are required to make this story a successful one, both on the experimental and on the theoretical side. In particular, the tremendous progress made by lattice QCD provided substantial input to the understanding of flavour violating processes, and further territories are waiting to be explored.

Instead of giving an exhaustive review, this article aimed to highlight a few important directions where recent lattice QCD results have had a significant impact on the understanding of flavour violating processes both in and beyond the SM. In the same spirit, let me close by collecting my top 5 wishes to the lattice QCD community for the coming years.
\begin{itemize}
\item In order to obtain more precise values for the CKM elements measured in tree-level decays, improved lattice QCD results for the {form factors entering the determinations of $|V_{ub}|$ and $|V_{cb}|$} from semileptonic decays are necessary. Their values are a crucial ingredient to make precise SM predictions for rare processes like $\eps_K$, $K\to\pi\nu\bar\nu$ and $B_{s,d}\to\mu^+\mu^-$ that are very sensitive to NP contributions.
\item To fully exploit the NP discovery potential of present $B$ physics experiments, precise lattice QCD results for the {$B_{d,s}$ mixing parameters and decay constants} are mandatory, in order to maximise their NP sensitivity. Further improvements are particularly required in light of the slight tension of the $B_{s,d}$ mixing data with their most recent SM prediction, and in light of the discovery of the decay $B_s\to\mu^+\mu^-$ whose measurement will be improved quickly.
\item The recent SM prediction of the ratio $\epe$, using the first lattice QCD results for the relevant $K\to\pi\pi$ amplitudes uncovered an intriguing discrepancy with the data. Improved lattice calculations for the hadronic matrix elements in question, as well as an independent confirmation by a different lattice collaboration, are badly awaited in order to confirm (or resolve) the current tension.
\item The various tensions observed in semileptonic $b\to s$ transitions received a lot of attention, in particular since they follow a consistent pattern in terms of a possible NP explanation. In order to disentangle NP effects from potentially underestimated non-perturbative QCD effects, more precise theory predictions are needed. While at the moment lattice QCD contributes here only to the understanding of the relevant hadronic form factors, it would be a real break-through if one day one could also access the non-factorisable corrections on the lattice.
\item Lastly, it would be very desirable to one day obtain solid lattice QCD predictions for the long-distance contributions entering neutral kaon and $D$ meson mixing, measured by the mass differences $\Delta M_K$ and $\Delta M_D$, respectively. Potential NP effects in these observables are currently obscured by the poor understanding of the SM long-distance dynamics. While lattice calculations of the long-distance dynamics entering $K^0-\bar K^0$ mixing are already under way, it will probably take a long time and require a lot of conceptual progress to access the long-distance dynamics of $D^0-\bar D^0$ mixing on the lattice. 
\end{itemize}

{\bf Acknowledgements. } I would like to thank the organisers of the Lattice 2016 conference for the invitation to present a talk in such a stimulating environment. 
Further I am grateful to Andrzej~J.~Buras for a careful reading and useful comments on this article.

\end{document}